\documentclass[aps,prl,reprint,showpacs,preprintnumbers,nofootinbib,superscriptaddress]{revtex4-1}
\usepackage{slashed}
\usepackage{bm} 
\usepackage{graphicx,epsf}
\usepackage{latexsym,amssymb,amsmath,float,url,mathrsfs}
\usepackage{soul}
\usepackage{natbib}
\usepackage{hyperref}
\usepackage{subfigure}

\newcommand{\gsim}{\mathrel{\vcenter{\hbox{$>$}\nointerlineskip\hbox{$\sim$}}}}

\newcommand{\beq}[1]{\begin{equation}\label{#1}}
\newcommand{\eeq}{\end{equation}}
\newcommand{\bea}{\begin{eqnarray}} 
\newcommand{\eea}{\end{eqnarray}}
\newcommand{\ba}{\begin{array}}
\newcommand{\ea}{\end{array}}

\newcommand{\dmsq}{\delta m^2_\eta}
\newcommand{\vev}{v_{\rm EW}}

\input{epsf}

\begin{document}

\title{Dark matter at the LHC: EFTs and gauge invariance}

\author{Nicole F.\ Bell}
\affiliation{ARC Centre of Excellence for Particle Physics at the Terascale \\
School of Physics, The University of Melbourne, Victoria 3010, Australia}

\author{Yi Cai}
\affiliation{ARC Centre of Excellence for Particle Physics at the Terascale \\
School of Physics, The University of Melbourne, Victoria 3010, Australia}

\author{James B.\ Dent}
\affiliation{Department of Physics, University of Louisiana at Lafayette, Lafayette, LA 70504-4210, USA}

\author{Rebecca K.\ Leane}
\affiliation{ARC Centre of Excellence for Particle Physics at the Terascale \\
School of Physics, The University of Melbourne, Victoria 3010, Australia}

\author{Thomas J.\ Weiler}
\affiliation{Department of Physics and Astronomy, Vanderbilt University, Nashville, TN 37235, USA}

\date{March 26, 2015}

\preprint{}

\begin{abstract}
Effective field theory (EFT) formulations of dark matter
interactions have proven to be a convenient and popular way to
quantify LHC bounds on dark matter.  However, some of the
non-renormalizable EFT operators considered do not respect the gauge
symmetries of the Standard Model.  We carefully discuss under what
circumstances such operators can arise, and outline potential issues
in their interpretation and application.
\end{abstract}

\pacs{95.35.+d, 12.15.Lk}


\maketitle


\noindent
\textbf{\textit{Introduction.}} \\ 
Despite overwhelming evidence that dark matter (DM) is the dominant form of
matter in the universe, we remain ignorant about its fundamental
nature.  An appealing class of DM candidates which enjoy considerable
theoretical motivation are weakly interacting massive particles
(WIMPs): heavy particles with weak-scale mass and interaction strengths.
Indeed, the production of WIMP-type particles at the LHC is now one of
the foremost goals of the particle physics community.

Given the large number of WIMP-type theories, it is desirable to
express the DM interactions in a model-independent manner.  This can
be achieved with an effective field theory (EFT) framework, in which a
set of non-renormalizable effective operators is used to parametrise
the interaction of a pair of DM particles with Standard Model (SM)
particles.  The EFT operators would be obtained as a low energy
approximation to a renormalizable theory by integrating out the
particle(s) that mediates the interaction.  A standard set of
operators have been listed in Refs.\cite{Goodman:2010yf,Goodman:2010ku} (see also~\cite{Duch:2014xda}). 

For fermionic dark matter $\chi$ interacting with SM fermions $f$,
these operators take the form:
\begin{equation}
\frac{1}{\Lambda^2}\left( \overline{\chi} \Gamma_\chi \chi \right) \left( \overline{f} \Gamma_f f\right) \; ,
\end{equation}
where $\Lambda$ has dimensions of mass and is related to the mass $M$
and coupling constants $g_i$ of a heavy mediator as
$\Lambda=M/\sqrt{g_1g_2}$, and $\Gamma_{\chi,f}$ are various Gamma matrices.  

While the EFT description is very useful at low energies, such as
those relevant for direct detection, it now well appreciated that the
EFT approach may be unsuitable at LHC energies.  Specifically, if the
momentum transfer in a process is comparable to or larger than the
mass of the mediator, the EFT will not provide an accurate description
of the underlying physics.  Many recent papers have attempted to
quantify the point at which an EFT description is no longer
valid~\cite{Busoni:2013lha,Busoni:2014sya,Busoni:2014haa} or have
proposed the use of simplified models as an alternative framework for
undertaking DM searches at colliders~\cite{Abdallah:2014hon,Buckley:2014fba,Alves:2011wf,Alwall:2008ag}.

Here we make a more subtle point: if an EFT operator does not respect
the weak gauge symmetries of the SM, it may be invalid at energies
comparable to the electroweak scale, $\vev\approx 246$~GeV, rather than the
energy scale of new physics, $\Lambda$.  For example, if we attempt to
use electroweak gauge symmetry violating operators at LHC energies,
serious difficulties can be encountered soon above the EW scale, such
as the bad high energy behaviour of cross sections.  An example is the
well-known unitary violation rising as $s/(4\,m_W^2)$ in
$SU(2)_L$ non-invariant $W W$ scattering, due to the longitudinal
$W$~modes induced by the symmetry breaking.  In the SM, the violations
are removed by an internal Higgs particle, but internal fields are
``integrated out'' in the EFT formalism.  Thus, the limit of validity
for the operator is the weak scale if any internal Higgs or $W$ or $Z$
particle is present in the Feynman diagram.  More generally, sacred
symmetries like the electroweak Ward identity can be violated, which
implies a weak-scale cutoff, as we explain later in this paper.


\bigskip
\noindent
\textbf{\textit{EFT Operators and Gauge Invariance.}} \\
The standard list of DM-SM effective operators~\cite{Goodman:2010ku}
contains several operators which violate the SM weak gauge symmetries. 
We argue that if an EFT operator does not respect the weak gauge
symmetries of the SM, it necessarily carries a pre-factor of the Higgs
vev to some power, a remnant of the $SU(2)_L$ scalar doublet
\beq{doubletPhi}
\Phi\equiv \left(
\begin{array}{c}
\phi^+ \\
\phi^0 = \frac{1}{\sqrt{2}} (H+\vev +i\Im\phi^0 )
\end{array}
\right)\,.
\eeq
Acting as an $SU(2)_L$ doublet, enough powers of $\Phi$ are required
to form an $SU(2)_L$-invariant operator.  The fields $\phi^\pm$ and
$\Im\phi$ are gauged away to become, in unitary gauge, the
longitudinal modes of the $W^\pm$ and $Z$.  So, in fact, it is the
real, neutral field $\frac{1}{\sqrt{2}} (H+\vev)$ whose $n^{th}$ power
appears in the operator.  Commonly, the $H$ part of the
expression is omitted, leaving just an implicit $\vev^n$ in the
coefficient.  Of course, the $\vev^n$ must come with a $\Lambda^{-n}$.
Omission of the $H$ part in the operator may ignore some
interesting phenomenology.  In this paper, we will also ignore the
$H$ contributions to operators, and focus on the operators
proportional to $(\vev/\Lambda)^n$.  Such terms in the coefficients of
$SU(2)$-violating operators clearly satisfy the criterion that as
$SU(2)$ symmetry is restored, $\vev\rightarrow 0$, the operator's
coefficient vanishes, and the operator decouples.~\footnote{In what follows, we will assume that there is but a single
  vev, $\vev$.  If there were further vevs, the good
  relation $m_W=\cos\theta_W\,m_Z$ requires the additional vevs to
  come from additional doublet fields, or to be be small if coming
  from non-doublet fields.  The vevs then add in quadrature to give
  $\left(2\,m_W/g_2\right)^2$.  Thus, any individual vev will
  offer an energy-scale below the SM vev.  In the sense that we will
  argue against larger energy-scales for effective operators, our
  assumption of a single EW vev is conservative.}

{\bf Scalar operator:}
Consider the scalar (or pseudo scalar) operators 
\begin{equation}
\frac{m_q}{\Lambda^3}\left( \overline{\chi} \chi \right) \left( \overline{q}  q\right)
=\frac{m_q}{\Lambda^3}\left( \overline{\chi} \chi \right) \left( \overline{q}_L  q_R + h.c. \right).
\end{equation}
This operator is clearly not $SU(2)_L$ invariant, as $\chi$ and $q_R$
are $SU(2)_L$ singlets, while ${q}_L$ is a component (either $u_L$ or
$d_L$) of the usual left-handed SM doublet, $Q_L$.
A coupling to the Higgs boson has been anticipated by the factor of $m_q$ in the
coefficient. Most authors invoke minimal flavor violation to
motivate this choice of normalization.  Although this $SU(2)_L$ violating
effective operator can be a good low energy description of new
physics, notice that its coefficient cannot be arbitrarily large as it
is controlled by the Higgs vev.  Although formally a dimension 6
operator, it is competitive only with dimension 7 operators, given
its $1/\Lambda^3$ normalization.

{\bf Vector operator:}   
Now consider vector (or axial vector) operators of the form
\begin{equation}
\frac{1}{\Lambda^2}\left( \overline{\chi} \gamma^\mu \chi \right) \left( \overline{q} \gamma_\mu  q\right)
=\frac{1}{\Lambda^2}\left( \overline{\chi} \gamma^\mu \chi \right) \left( \overline{q}_L \gamma_\mu q_L + \overline{q}_R \gamma_\mu q_R  \right)\,.
\label{vectorOp}
\end{equation}
These operators respect $SU(2)_L$ provided that the coefficients of
the $u_L$ and $d_L$ operators are equal\footnote{Isospin violating
operators, such as those invoked in \cite{Feng:2011vu,Feng:2013vod},
can obviously be crafted from the RH quark fields.}, 
Any $(\overline{u}_L \gamma_\mu u_L)$ operator that does not have a
matching $d_L$ term 
should be suppressed by two powers of $\vev/\Lambda$ (one for each unmatched $u_L$):
\begin{equation}
\frac{\vev^2}{\Lambda^4} (\overline\chi \gamma^\mu \chi)(\overline{u_L} \gamma_\mu u_L).
\end{equation}
Including the suppressed coefficient, this SU(2)-violating operator
competes with dimension 8 operators, i.e., while the $SU(2)$
conserving (axial)vector operators are dimension 6, $SU(2)$ violating
(axial)vector operators compete with subdominant, higher-order,
dimension 8 operators.



\bigskip
\noindent
\textbf{\textit{Mono-$W$ and $SU(2)_L$ invariance.}}\\
Issues arise if one tries to use gauge symmetry violating operators at
LHC energies.  For particular processes, the lack of gauge invariance
can manifest as a violation of unitarity in high energy scattering.
As an example of a problem encountered with an $SU(2)_L$ violating
EFT, consider the following operator:
\begin{equation}
\frac{1}{\Lambda^2}\left( \overline{\chi} \gamma^\mu \chi \right) \left( \overline{u} \gamma_\mu u + \xi \overline{d} \gamma_\mu d\right).
\label{eq:xi}
\end{equation}
This Lagrangian violates $SU(2)_L$, unless $\xi=1$.  
The case of unequal $u$ and $d$ couplings was considered in
Ref.~\cite{Bai:2012xg}, where a very strong constructive(destructive)
``interference effect'' was found for $\xi=-1 (+1)$, the
degree of which depends on the energy scale.
The analysis of Ref.~\cite{Bai:2012xg} was subsequently repeated by the
LHC experimental collaborations ATLAS~\cite{Aad:2013oja,ATLAS:2014wra}
and CMS~\cite{CMS:2013iea,Khachatryan:2014tva}.
We shall demonstrate that the large cross section enhancement for $\xi
\neq +1 $ is in fact due the production of longitudinally polarized $W$'s
as a result of breaking gauge invariance.

\begin{figure}[b]
\centering
  \subfigure[$\mathcal{M}_1$]{\includegraphics[width=0.4\columnwidth]{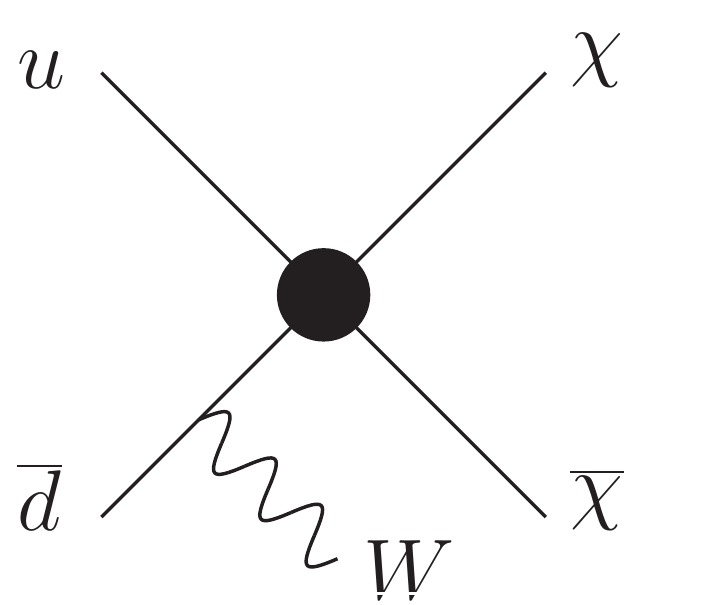}}
  \subfigure[$\mathcal{M}_2$]{\includegraphics[width=0.4\columnwidth]{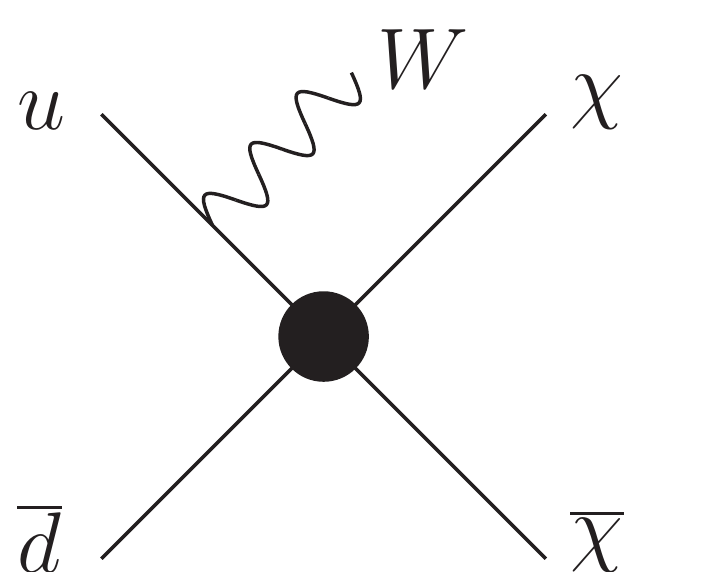}}
\caption{Contributions to the mono-$W$ process $ u(p_1)\overline{d}(p_2)  \rightarrow  \chi(k_1)\overline{\chi}(k_2)  W^+(q)$, in the effective field theory framework.}
\label{fig:monoWeft}
\end{figure}

At parton level, the mono-$W$ process is $u(p_1)\overline{d}(p_2)
\rightarrow \chi(k_1)\overline{\chi}(k_2) W^+(q)$.  The relevant
diagrams are given in Fig.\ref{fig:monoWeft}, and the corresponding
contributions to the amplitude $\mathcal{M} \equiv
\mathcal{M}^\alpha\epsilon^\lambda_\alpha(q) \equiv
(\mathcal{M}_1^\alpha +
\mathcal{M}_2^\alpha)\epsilon^\lambda_\alpha(q)$ are
\begin{eqnarray}
  \mathcal{M}_1^\alpha &=& \frac{1}{\Lambda^2}\left[\bar{v}(p_2)\gamma^\alpha\frac{-g_W}{\slashed{p_2}-\slashed{q}}\gamma^\mu \frac{P_L}{\sqrt{2}} u(p_1)\right]\left[\bar{u}(k_1)\gamma_\mu v(k_2)\right], 
\nonumber \\
\mathcal{M}_2^\alpha &=& \frac{\xi}{\Lambda^2}\left[\bar{v}(p_2)\gamma^\mu\frac{g_W}{\slashed{p_1}-\slashed{q}}\gamma^\alpha \frac{P_L}{\sqrt{2}} u(p_1)\right]\left[\bar{u}(k_1)\gamma_\mu v(k_2)\right], \;\;\;\;\;\;
\label{eq:amp}
\end{eqnarray}
where $g_W$ is the weak coupling constant, and
$\epsilon^\lambda_\alpha$ is the polarization vector of the $W$.  We
note that the $W$ longitudinal polarization vector at high energy is
\begin{equation}
\epsilon_\alpha^L = \frac{q_\alpha}{m_W} + \mathcal{O}\left(\frac{m_W}{E}\right) 
\sim \frac{\sqrt{s}}{m_W} .  
\end{equation}
Thus the high energy $W_L$ contribution to the usual polarization sum,
$\sum_\lambda\epsilon^\lambda_\alpha\epsilon^{\lambda\,*}_\beta=-g_{\alpha\beta}+\frac{q_\alpha q_\beta}{m_W^2}$,
is $\epsilon^L_\alpha \epsilon^{L\,*}_\beta \approx q_\alpha q_\beta/m_W^2 \sim s/m_W^2$.

We can verify that the sum of the two amplitudes of
Fig.\ref{fig:monoWeft} is not gauge invariant unless $\xi=1$, by
observing that the relevant Ward identity is not satisfied.
At high energy, the Goldstone boson equivalence theorem requires that
the amplitude for emission of a longitudinally polarized $W_L$ is
equivalent to that for the emission of the corresponding Goldstone
boson.  Since the Goldstone boson couples to quarks with strength
proportional to their mass, these terms are close to zero. (See
Ref.\cite{Ciafaloni:2010qr} for a similar discussion about the related
process $\chi\chi \rightarrow \nu e W$.)  
The Ward identity for the longitudinal $W$ at high energy therefore takes the form
\begin{equation}
\mathcal{M}^\alpha \epsilon^L_\alpha \approx \frac{q_\alpha}{m_W} \mathcal{M}^\alpha(q,...) = i \mathcal{M}(\phi^+(q))\simeq 0.
\end{equation}
%
For the sum of the mono-$W$ amplitudes of Fig.\ref{fig:monoWeft} we find
\begin{equation}
q_\alpha \mathcal{M}^\alpha
= \frac{g_W}{\Lambda^2}\left[\bar{v}(p_2)\left(1  - \xi \right) \gamma^\mu \frac{P_L}{\sqrt{2}} u(p_1)\right]
\left[\bar{u}(k_1)\gamma_\mu v(k_2)\right],
\end{equation}
which clearly vanishes only for $\xi=1$.

The ``interference effect" seen in the mono-W process is not truly due
to constructive/destructive interference as previously claimed, but is
just a manifestation of the fact that the breaking of electroweak
gauge-invariance has given rise to a $W_L$ component.  The increased
cross section for $\xi\neq 1$ is in fact due to unphysical terms that
grow like $s/m_W^2$, which originate from the $+q_\alpha q_\beta/m_W^2$
term in the polarization sum.  At high energy, these terms would grow
large enough to violate unitarity.  But even at lower energy, their
presence may be problematic.

To explicitly demonstrate this behaviour,
we now derive an analytic expression for the parton-level mono-$W$
process $\overline{d} u \rightarrow \overline{\chi} \chi W^+$.  We
work in the center-of-mass frame, and follow the phase space
parametrization described in Section V of Ref.\cite{Schmeier:2013kda}.
We define $\theta$ to be the angle of the $W$ w.r.t. the beam line
and $x=2E_W/\sqrt{s}$, where $\sqrt{s}$ is the total invariant mass.
For simplicity we take $m_\chi=0$ (the cross section will be
approximately independent of $m_\chi$ for $m^2_\chi \ll s$). 
We include a factor of $1/3$ from averaging over initial state quark colors.

For $\xi=1$ the differential cross section is well behaved and is given by 
\begin{widetext}
\begin{eqnarray}
\label{xi1cross}
\left. \frac{d^2\sigma}{dxd\cos\theta } \right|_{\xi=1}
=\frac{A}{3^2 2^8\pi ^3 \Lambda ^4 \left(s^2 x^2 \sin ^2 \theta +  2 s m_W^2 \left(\cos  \left(2 \theta \right)-2 x+1\right)+4 m_W^4
\right)^2},
\end{eqnarray}
where
\begin{align*}
A&= s^2 g_W^2 \sqrt{x^2- \frac{4 m_W^2}{s}} \left( 1-x + \frac{m_W^2}{s}\right) \Big[
s^3 x^2 \sin ^2 \theta  \left( \cos  \left(2 \theta \right)x^2+3 x^2-8 x+8\right)  \\
& 
 + 2 s^2 m_W^2 \left( \cos  \left(4 \theta \right) x^2 +2 \cos  \left(2 \theta \right)\left(x^3-x^2-4 x+4\right)  -2 x^3+17 x^2-24 x+8\right)  \\
&  
-4 s m_W^4 \left(\cos  \left(4 \theta \right)+ \cos \left(2 \theta \right)\left(x^2+4 x-8\right)-x^2+20 x-17\right) +16 m_W^6 \left(\cos  \left(2 \theta \right)+3 \right) \Big],
\end{align*}
If we take the limit $m_W\rightarrow0$, remove the color factor $1/3$,
and replace $g_W/\sqrt{2}$ with the electron charge $e$, we find Eq.~\ref{xi1cross} reproduces the cross section for the $e^+e^-\rightarrow \overline{\chi}\chi\gamma$ monophoton process reported by Ref.\cite{Schmeier:2013kda,
  Chae:2012bq} for $m_\chi=0$ and unpolarized $e^+e^-$ beams.  This provides a useful check for our more complicated  mono-$W$ calculation.

For $\xi \neq 1$, however, the cross section is not well behaved at
high energy.  The $+q_\alpha q_\beta/m_W^2$ term in the polarization
sum
%
%
contributes to the cross section a term
\begin{equation}
\label{qqsigma}
\left. \frac{d^2\sigma}{dxd\cos\theta } \right|_{q_\alpha q_\beta/m_W^2}
=\frac{(\xi -1)^2 s^2 g_W^2 \sqrt{x^2-\frac{4 m_W^2}{s}} \left(
2 x^2 \sin^2 \theta - 16 x+16 +
\frac{4 m_W^2}{s} \left(\cos  \left(2 \theta\right)+3\right)
\right)}{3^2 2^{13}\pi ^3 \Lambda ^4 m_W^2},
\end{equation}
\end{widetext}
which violates unitarity when $s \gg m^2_W$.

\begin{figure}[t]
\centering
\subfigure{\includegraphics[width=1.0\columnwidth]{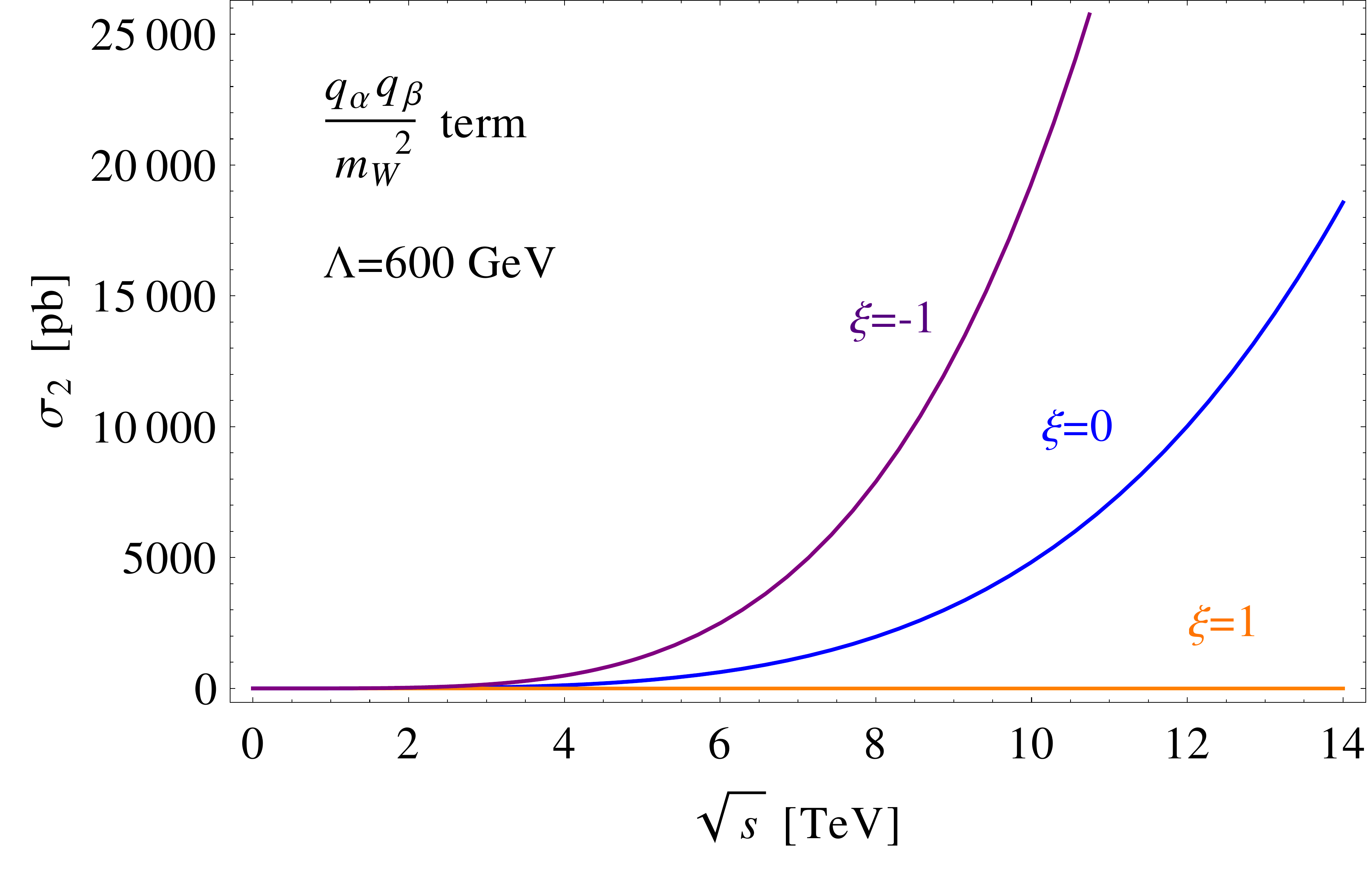}}
\subfigure{\includegraphics[width=1.0\columnwidth]{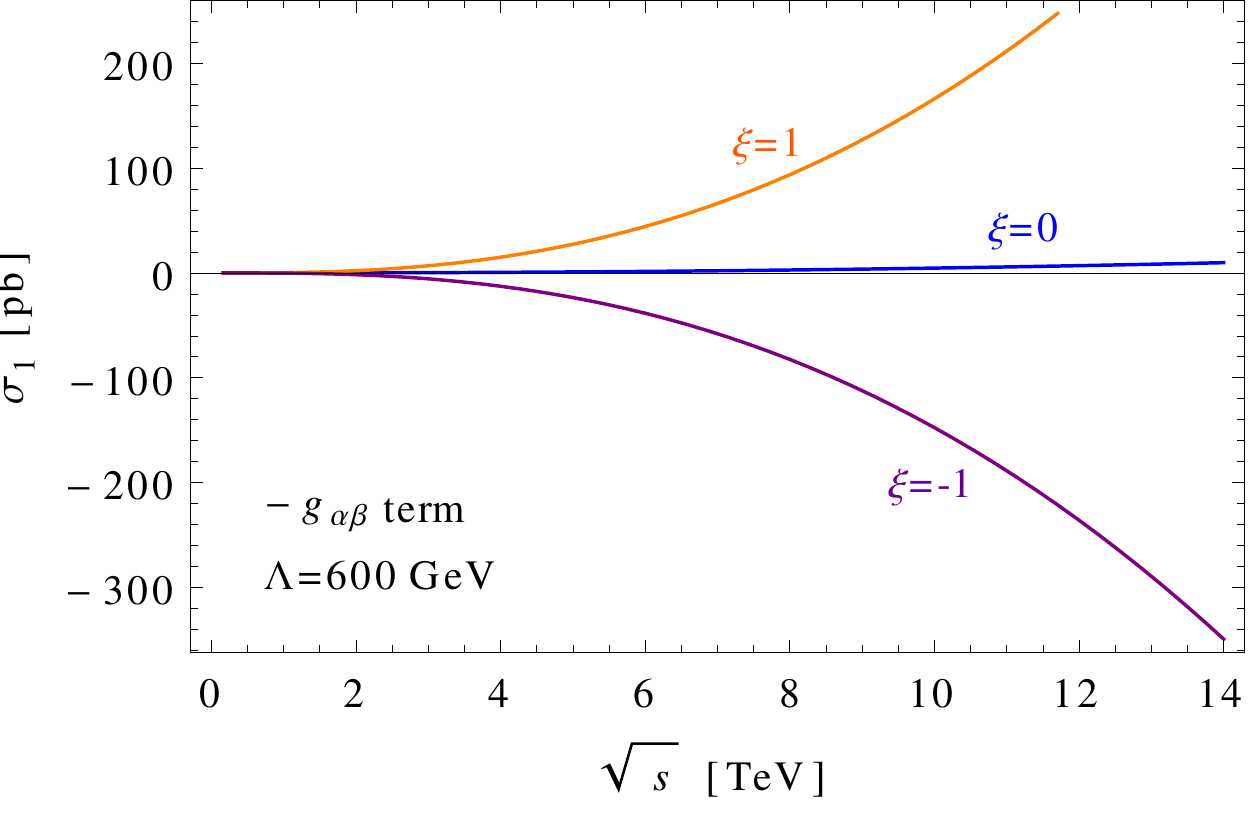}}
\caption{Total parton-level cross sections as a function of energy,
  for $\Lambda=600$~GeV and particular choices of $\xi$.  Upper:
  contribution from the $+q_\alpha q_\beta/m_W^2$ term in the
  polarization sum. The cross section scales simply as $(1-\xi)^2$.
  Lower: contribution of the $-g_{\alpha\beta}$ term in the
  polarization sum.  At LHC energies the $q_\alpha q_\beta$ terms
  dominates unless $\xi\simeq 1$.  (Notice the differing vertical
  scales between the upper and lower panels.) }
\label{fig:new}
\end{figure}

\begin{figure}[t]
\vspace{-2mm}
\includegraphics[width=1.0\columnwidth]{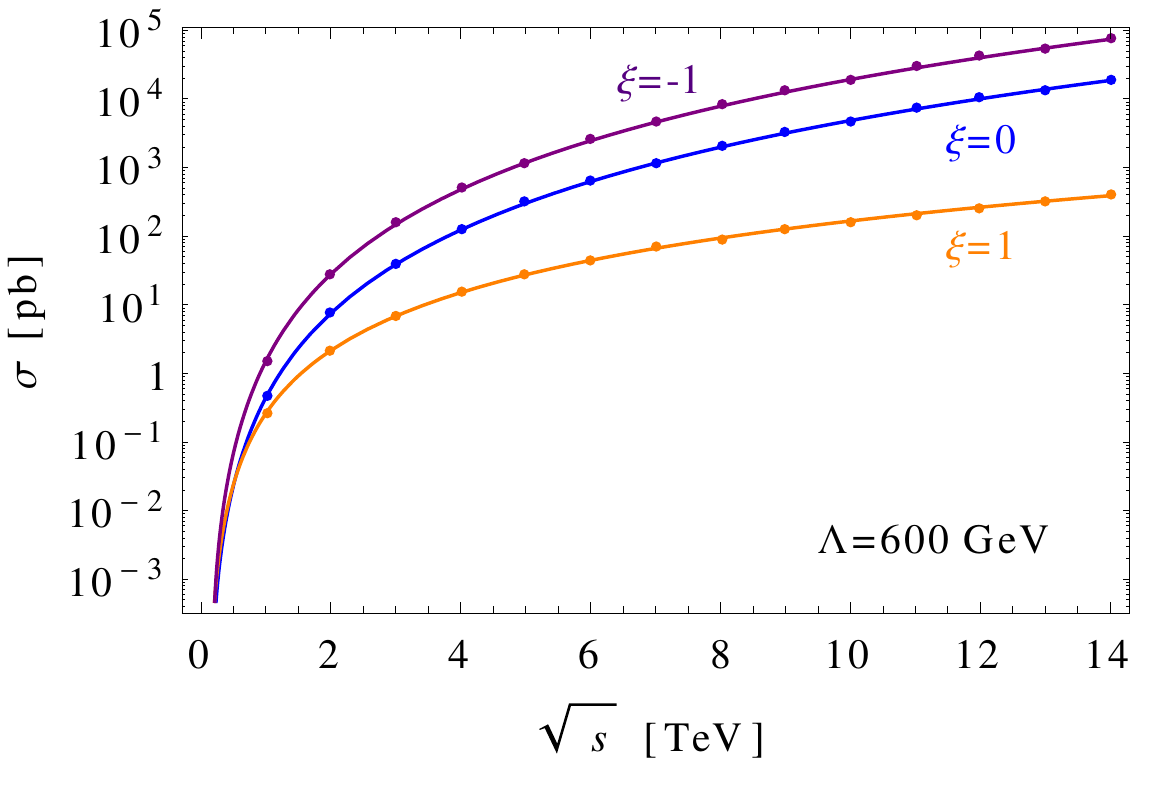}
\caption{Total parton-level cross sections for $\Lambda=600$~GeV, for
  particular choices of $\xi$. Solid lines are the analytic
  calculation and dots are the MadGraph calculation.  }
\label{fig:old}
\end{figure}

The total cross sections, for $m_\chi=0$, are plotted in
Figs(\ref{fig:new},\ref{fig:old}) as a function of $\sqrt{s}$. We also
calculate the cross sections in MadGraph \cite{Alwall:2011uj}, and
find the results agree.  For brevity of notation, we have defined
$\sigma _1$ and $\sigma_2$ to be the contributions to the cross
section from the $-g_{\alpha\beta}$ and $+q_\alpha q_\beta/m_W^2$ terms
in the polarization sum, respectively.  The $\xi =0, -1$ cross
sections grow faster with $\sqrt{s}$ than for $\xi=1$.  At LHC
energies the cross sections are already dominated by the unphysical
terms arising from the longitudinal polarization, unless $\xi\simeq
1$.


\bigskip
\noindent
\textbf{\textit{From Renormalizable Models to EFTs.}}\\
Let us now consider a renormalizable, gauge invariant, model of DM
interactions, and examine the way in which unequal couplings to $u$ and
$d$ quarks can be obtained.  Consider the case where
$\overline{q}q\rightarrow\overline\chi \chi$ is mediated by the exchange of a
$t$-channel scalar.  The Lagrangian is given by
 \begin{eqnarray}
\mathcal{L}_\text{int}&=&f \overline{Q_L} \eta \chi_R+h.c
\nonumber\\ 
&=&f_{ud}\left(\eta_u\overline{u}_L+\eta_d\overline{d}_L\right)\chi_R+h.c.,\label{eq:tchannel}
\end{eqnarray}
where $Q_L = (u_L,d_L)^T$ is the quark doublet,
$\eta=(\eta_u,\eta_d)^T \sim (3,2,1/3)$ is a scalar field that
transforms under the SM gauge group like $Q_L$, and $f$ is a coupling
constant.  Such couplings are present in supersymmetric (SUSY) models,
with $\chi$ identified as a neutralino and $\eta$ a squark doublet,
and have been considered as a simplified model for DM interactions in
Refs.~{\cite{Difranzo:2013, Berger:2013, Wang:2014, Chang:2014}}.

If we take the EFT limit, assuming the $\eta$ are very heavy, the
lowest order operators are of dimension 6:
\begin{eqnarray}
\frac{1}{\Lambda_u^2}(\overline{u}\Gamma u)(\overline\chi \Gamma \chi)
\quad\textrm{   and   }\quad 
\frac{1}{\Lambda_d^2}(\overline{d}\Gamma d)(\overline\chi
\Gamma \chi),
\end{eqnarray}
 where the suppression scales are $\Lambda_{u,d} \propto
 m_{\eta_{u,d}}/f$. The relevant Lorentz structure $\Gamma$ is a sum
 of vector and axial vector terms as can be seen by Fierz transforming
 the t-channel matrix elements obtained from Eq.\ref{eq:tchannel} to
 s-channel form~\cite{Bell:2010ei}.

The strength of DM interactions with $u$ and $d$ quarks can differ
if the masses of $\eta_u$ and $\eta_d$ are non-degenerate.  However,
given that $(\eta_u,\eta_d)$ form an electroweak doublet, their mass
splitting must be controlled by $\vev$.  The relevant terms
in the scalar potential are~\cite{Garny:2011cj} 
\begin{eqnarray}
V &=& 
m_1^2 (\Phi^\dagger \Phi) +
\frac{1}{2}\lambda_1 (\Phi^\dagger \Phi)^2 +
m_2^2 (\eta^\dagger \eta) +
\frac{1}{2}\lambda_2 (\eta^\dagger \eta)^2   \nonumber \\
&+& \lambda_3 (\Phi^\dagger \Phi) (\eta^\dagger \eta) +
\lambda_4 (\Phi^\dagger \eta) (\eta^\dagger \Phi).
\end{eqnarray}
If $m_1^2<0$ and $m_2^2>0$, the SM Higgs doublet obtains a non-zero
vev, while the $\eta$ does not.  The presence of $\lambda_4$ splits
the $\eta$ masses as
\begin{eqnarray}
&& m^2_{\eta_d}=m_2^2 + (\lambda_3 +\lambda_4)v_{\textrm EW}^2, \\
&& m^2_{\eta_u}=m_2^2 + \lambda_3 v_{\textrm EW}^2\,,
\end{eqnarray}
implying that $\dmsq \equiv m^2_{\eta_d} - m^2_{\eta_u} = \lambda_4\,v_{\textrm EW}^2$.
%
%
Note that while we have engineered unequal scalar masses, and thus
unequal DM couplings to $u$ and $d$ quarks, we do not have complete
freedom.  The parameter $\xi$ of Eq.\ref{eq:xi} is given by $\xi=1/(1+
\dmsq/\Lambda^2) = 1/(1+\lambda_4\,v_{EW}^2/\Lambda^2$).  For $\Lambda
\gsim 1$~TeV and a perturbative value for $\lambda_4$, $\xi$ will not
deviate far from 1.  
(Negative $\xi$ can not be obtained from our renormalizable model.)
Furthermore, it is clear that $SU(2)_L$ violating effects enter the EFT at order
$v^2_{EW}/\Lambda^4$, i.e., the same order in $\Lambda$ as a dimension
8 operator.  

\begin{figure*}[ht]
\subfigure[]{\includegraphics[width=0.5\columnwidth]{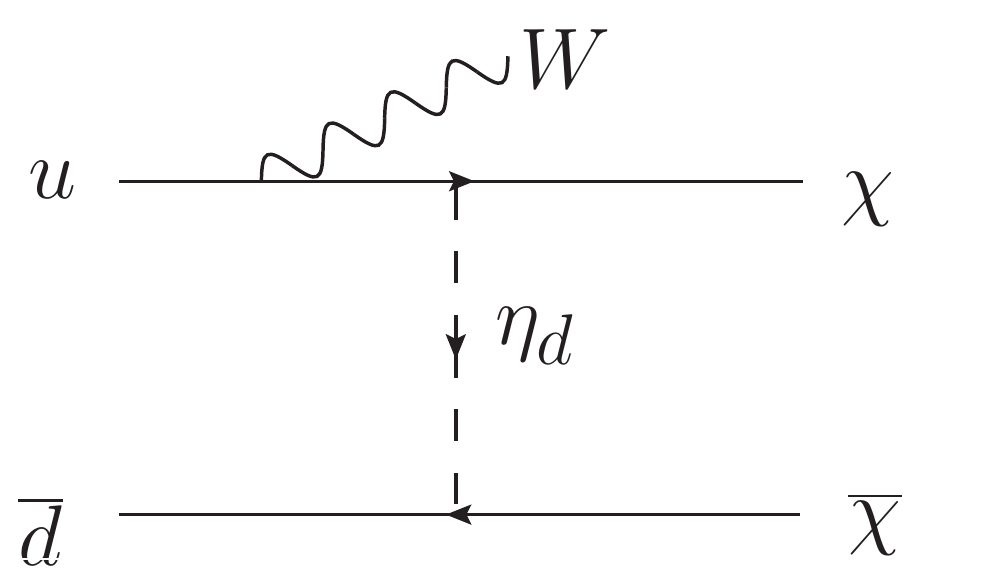}}
\subfigure[]{\includegraphics[width=0.5\columnwidth]{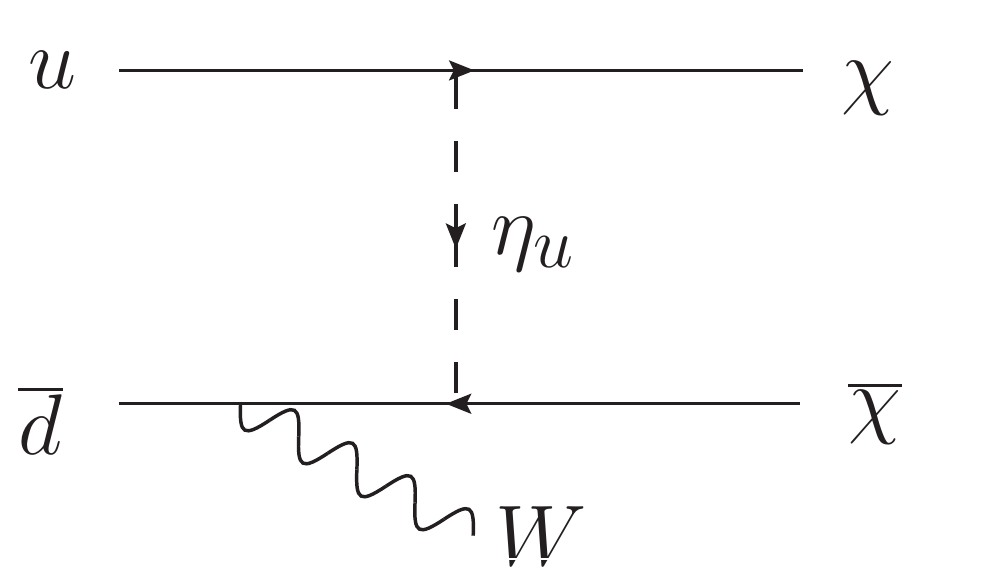}}
\subfigure[]{\includegraphics[width=0.5\columnwidth]{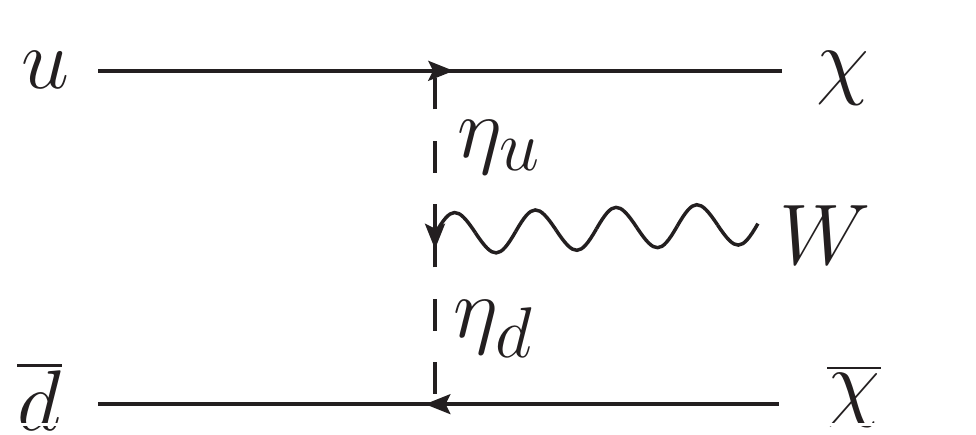}}
\caption{Contributions to the mono-$W$ process $ u(p_1)\overline{d}(p_2)  \rightarrow  \chi(k_1)\overline{\chi}(k_2)  W^+(q)$, in a UV complete model.}
\label{fig:diags}
\end{figure*}

In the renormalizable theory, the mono-$W$ process proceeds via the
gauge invariant set of diagrams in Fig.\ref{fig:diags}~\footnote{In the good
  EW $SU(2)$ limit, the $\eta_u$ and $\eta_d$ are mass degenerate, and
  the massless $W^\pm$ emitted in diagram (c) establishes the validity
  of the EW Ward identity~\cite{Bell:2011if,Ciafaloni:2011sa}.  When
  EW $SU(2)$ is broken, the $\eta_u$ and $\eta_d$ masses are split,
  and the new massive-$W$ longitudinal mode must restore the EW Ward
  identity by coupling to the $\eta$ proportional to
  $\dmsq$~\cite{Garny:2011cj}.  This argument provides an
  interpretation of the result found earlier in~\cite{Garny:2011cj}
  that the internal longitudinal mode couples proportional to $\dmsq$.
  In fact, in~\cite{Garny:2011cj} it was shown that this longitudinal
  $W$ mode will dominate the $W$ emission probability for some range
  of model
  parameters.}~\cite{Bell:2011if,Ciafaloni:2011sa,Bell:2012rg}.  
In the EFT limit, the diagrams in Fig.\ref{fig:diags}(a) and (b) map
onto those in Fig.\ref{fig:monoWeft}(a) and (b) respectively.  The
diagram in Fig.\ref{fig:diags}(c), in which the $W$ is radiated from
the $\eta$, is suppressed by an additional heavy scalar propagator,
and hence appears subdominant to the ISR diagrams.  It enters the EFT
as a dimension 8 operator, contributing on an equal footing with the
$SU(2)$ violating contributions of diagrams (a) and
(b)~{\cite{DeSimone:2013gj}.  Finally, note that in the renormalizable
  theory, in the high energy limit, $W_L$ production arises solely
  from the amplitude of Fig.\ref{fig:diags}(c),
and only when $\dmsq\neq 0$.


\medskip
\noindent
\textbf{\textit{Conclusion.}}\\
An important observation of Ref.\cite{Bai:2012xg} is that, of the
mono-$X$ processes, the mono-$W$ is unique in its ability to probe
different DM couplings to $u$ and $d$ quarks.  This important insight
is correct.  However, we have argued that the size of any $SU(2)_L$
violating difference of the $u$ and $d$ quark couplings must be
protected by the EW scale, and therefore cannot be arbitrarily large.
$SU(2)_L$ violating operators can be obtained by integrating out the
SM Higgs or by including Higgs vev insertions.
Therefore, they should have coefficients suppressed by powers of
$(\vev/\Lambda)$ or $(m_{\rm fermion}/\Lambda)$ and thus are of higher
order in $1/\Lambda$ than they would naively appear.
To include $SU(2)$ violating effects in a way that is self consistent
and properly respects the EW Ward identity, one should use a
renormalizable, gauge invariant, model rather than an EFT,
to avoid spurious $W_L$ contributions.
These observations will be an important guide to the LHC
collaborations in the interpretation of their
current~\cite{Aad:2013oja,ATLAS:2014wra,CMS:2013iea,Khachatryan:2014tva}
and forthcoming mono-$W$ dark matter search results, and to theorists
constructing EFTs.
%


\medskip
\noindent
\textbf{\textit{Acknowledgements.}}
NFB and YC were supported by the Australian
Research Council, RKL by the Commonwealth of
Australia and TJW by DOE grant DE-SC0010504.

\vspace{-4mm}


\bibliography{gauge_mar_26}


\end{document}